\documentclass[pre,twocolumn,superscriptaddress,floatfix]{revtex4-1}  

\usepackage[paperwidth=210mm,paperheight=297mm,centering,hmargin=2cm,vmargin=2.5cm]{geometry}

\usepackage[pdftex]{graphicx}
\usepackage{amsmath,amsfonts,amssymb}
\usepackage{natbib}
\usepackage{MnSymbol}
\usepackage{csquotes}

\usepackage[]{algorithm}
\usepackage{algorithmic}

\usepackage{hyperref}
\usepackage{xcolor}

\definecolor{bluemoi}{rgb}{0.25,0.50 ,0.75} 

\hypersetup{
  bookmarksopen=true,
  pdftitle=" ",
  pdfauthor=" ", 
  pdfsubject=" ", 
  pdftoolbar=true, 
  pdfmenubar=true, 
  pdfhighlight=/O, 
  colorlinks=true, 
  pdfpagemode=UseNone, 
  pdfpagelayout=SinglePage, 
  pdffitwindow=true, 
  linkcolor=bluemoi, 
  citecolor=bluemoi, 
  urlcolor=bluemoi 
}

\makeatletter
\renewcommand{\figurename}{\sf \textbf{Figure}}
\renewcommand{\thefigure}{\arabic{figure}}
\renewcommand{\fnum@figure}{\sf\textbf{\figurename}~\textbf{\thefigure}}
\renewcommand{\tablename}{\sf\textbf{Table}}
\renewcommand{\thetable}{\arabic{table}}
\renewcommand{\fnum@table}{\sf\textbf{\tablename}~\textbf{\thetable}}
\makeatother

\begin{document}

\title{On the importance of trip destination\\ for modeling individual human mobility patterns} 

\author{Maxime Lenormand}
\affiliation{TETIS, Univ Montpellier, AgroParisTech, Cirad, CNRS, INRAE, Montpellier, France}

\author{Juan Murillo Arias}
\affiliation{BBVA Data \& Analytics, Avenida de Burgos 16D, 28036 Madrid, Spain}

\author{Maxi San Miguel}
\affiliation{Instituto de F\'isica Interdisciplinar y Sistemas Complejos IFISC (CSIC-UIB), Campus UIB, 07122 Palma de Mallorca, Spain}

\author{Jos\'e J. Ramasco}
\affiliation{Instituto de F\'isica Interdisciplinar y Sistemas Complejos IFISC (CSIC-UIB), Campus UIB, 07122 Palma de Mallorca, Spain}

\begin{abstract}
Getting insights on human mobility patterns and being able to reproduce them accurately is of the utmost importance in a wide range of applications from public health, to transport and urban planning. Still the relationship between the effort individuals will invest in a trip and its purpose importance is not taken into account in the individual mobility models that can be found in the recent literature. Here, we address this issue by introducing a model hypothesizing a relation between the importance of a trip and the distance traveled. In most practical cases, quantifying such importance is undoable. We overcome this difficulty by focusing on shopping trips (for which we have empirical data) and by taking the price of items as a proxy. Our model is able to reproduce the long-tailed distribution in travel distances empirically observed and to explain the scaling relationship between distance traveled and item value found in the data.
\end{abstract}

\maketitle

\section*{Introduction}

Individual human mobility is a complex phenomenon, involving various mechanisms interacting at different spatial and temporal scales. These dynamics are the product of individual behaviors, governed by decisions that may depend on multiple contextual factors such as economic resources, geography, culture, norms, habits or life experiences. However, beneath this apparent complexity lies remarkable temporal and spatial regularities in the way people travel and interact with their environment \cite{Barbosa2018}. Results obtained in several studies based on dollar-bill tracking \cite{Brockmann2006}, mobile phone data \cite{Gonzalez2008}, Twitter data \cite{Hawelka2014,Lenormand2015b}, Foursquare data \cite{Cheng2011} and GPS data \cite{Raichlen2014} suggest that the distance $\Delta_r$ between two consecutive locations follows an heavy-tailed distribution well approximated by a Pareto function $P(\Delta_r) \sim \Delta_r^{-(1+\alpha)}$ with $0 < \alpha \leq 1$. It has also been shown that individuals tend to be attracted by popular places \cite{Roth2011,Hasan2013} and to return to previously visited locations, thus increasing the predictability of individual human movements \cite{Song2010a} and allowing the identification of most visited locations as well as the characterization of daily commuting patterns \cite{Schneider2013}. Individual human mobility patterns are also strongly influenced by geographical constraints \cite{Kang2012} but also by individuals' socio-economic status \cite{Lotero2014,Lenormand2015a,Gauvin2019} and social network \cite{Wang2011,Grabowicz2014,Picornell2015,Toole2015}.

Based on these empirical observations, several models have been proposed for modeling individual human mobility patterns. The simplest type describes human traveling using L{\'e}vy Flights and Continuous Time Random Walks \cite{Brockmann2006,Rhee2008}. These models give accurate predictions but fail at reproducing some features such as the individuals' tendency for revisiting locations \cite{Gonzalez2008,Rhee2008,Song2010b}. In \cite{Song2010b}, the authors propose a new model considering two generic mechanisms: exploration and preferential return, to decide whether an individual will visit a new place or a previously visited one as his/her next displacement. Going further in this direction, several models have been proposed to take into account diverse contextual factors such as the social context, urban geography and/or type and popularity of locations \cite{Lee2009,Kang2012,Hasan2013,Schneider2013}.

Nonetheless, most of these models focus on stationary (long-term) mobility, and, most importantly, they do not take into account the characteristics of the destination such as the travel purpose and its importance for the individual. Indeed, one can assume that an individual is not willing to invest the same amount of time or money, more generically, the same effort or amount of \enquote{energy} into a travel according to the value attached to the destination/objective of this travel. A basic trip purpose is the displacement between home and work, which have been collected in censuses for decades (in the US, for instance, since 1990). The introduction of new GPS-based technologies have permitted to explore other trip purposes since the early 2000's \cite{wolf2001,bohte2009}. Even though the relationship between trip cost and destination importance has been postulated in transport economy, and more recently in ecology, with the use of travel cost methods to assess the value of a natural sites based on the time and travel cost expenses that people spent to visit this site \cite{Parsons2003,Butterfield2016}, without adequate empirical data sets to explicitly assess the \enquote{value} of a destination this feature is rarely modeled at an individual scale.

\begin{figure*}
	\centering
	\includegraphics[width=15cm]{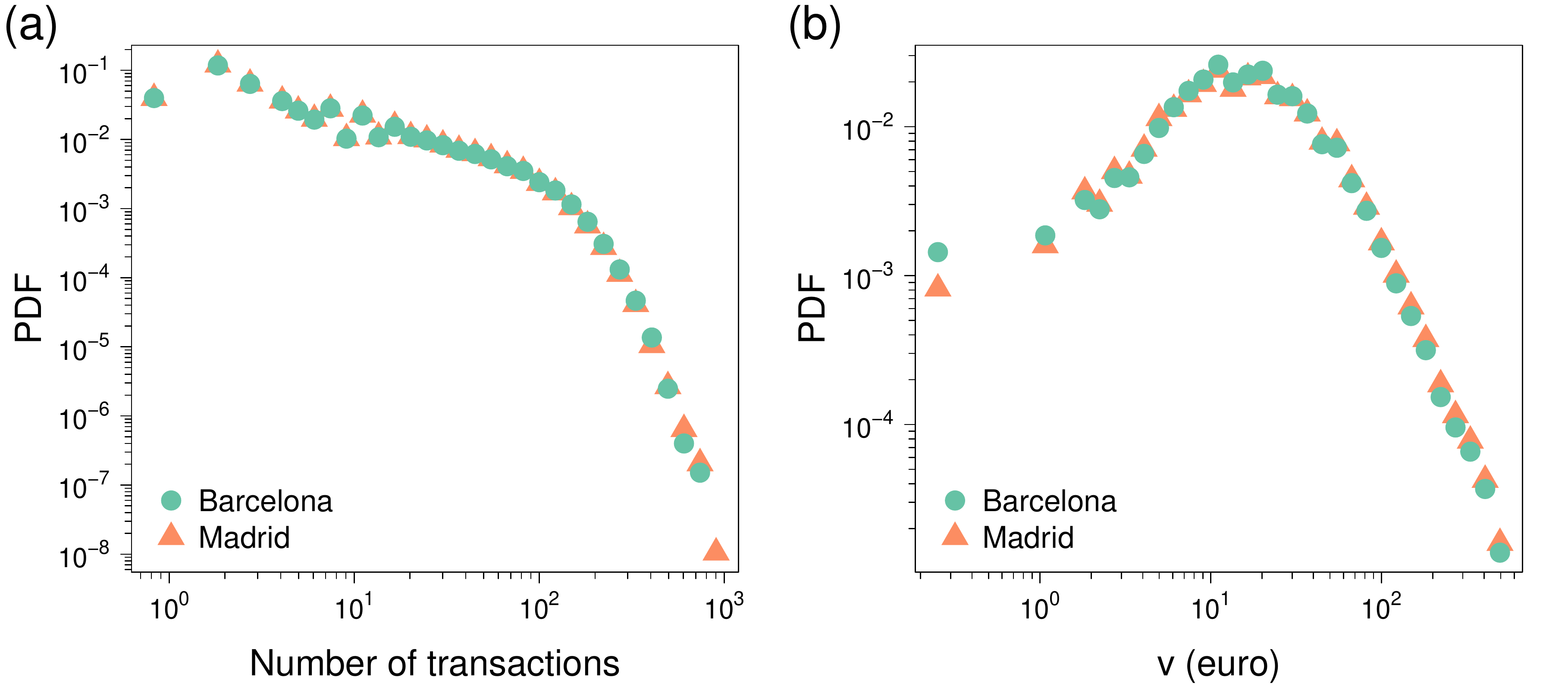}
	\caption{\textbf{Probability density function of the number of transactions per user in 2011 (a) and the amount of money spent per transaction (b) in Barcelona (green dots) and Madrid (orange triangles).} \label{Fig1}}
\end{figure*}

The purpose of this work is to understand the displacement distribution generated by a process in which the trips have a clear purpose and, therefore, an associated objective or subjective value $v$. The main assumption, straightforwardly checked in the data, is that the trips' length, $d$, tends to increase with $v$. We start by presenting a shopping mobility dataset that we use in the analysis and in which we can assign an objective meaning to $v$ as the price of the purchased items. This dataset contains information on bank card transactions made in the provinces of Barcelona and Madrid. Inspired by search processes for wild food resources in natural environment \cite{Viswanathan1996,Viswanathan1999,Viswanathan2010,Raichlen2014}, we introduce a human individual mobility model taking into account the value given to the trip destination through a parameter $p$, accounting for the probability of stoping or satisfying a search and that decreases when the value of $v$ increases. The model generates trip length distributions that mimic the empirical ones and it is able to explain the observed scaling relations from the data. 

\section*{Material and Methods}

\subsection*{Data}

To explore the relationship between travel cost $d$ and the importance given to its destination $v$, we analyze a credit card dataset containing information about $35$ million bank card transactions made by card holders (hereafter called users) of the Banco Bilbao Vizcaya Argentaria (BBVA) in the province of Barcelona and Madrid in $2011$. Each transaction is characterized by its amount (in euro currency) and a timestamp. Each transaction is also linked to a user and a business using anonymized user and business IDs. Users are identified with an anonymized user ID and their postcode of residence. In the same way, businesses are identified with an anonymized business ID, a business category (accommodation, automotive industry, bars and restaurants, etc.) and the geographical coordinates of the credit card terminal (see Table S2 in Supplementary Information (SI) for a full list of the selected business categories). The mobility habits and the representativeness of the BBVA credit card users in Barcelona and Madrid have already been investigated in
\cite{Lenormand2015a,Louail2017}. Here, we filtered out users with an average number of transactions per day higher than three (see the SI for more details). Only credit card payments whose amount was inferior to $500$ euros have been considered. Table \ref{Tab1} presents the final number of users, businesses and transactions analyzed in this study.

\begin{table}[!h]
	\caption{\textbf{Number of users, businesses and transactions in both case studies.} The number of postcodes and inhabitants and the surface area of the two provinces are also displayed.}
	\label{Tab1}
	\begin{center}
		\begin{tabular}{lcc}
			\hline
			\centering Statistics   & Barcelona  &  Madrid       \\
			\hline
			Number of postcodes    & 364         &  268           \\
			Number of inhabitants  & 5,540,925   &  6,489,680     \\
			Area (km$^2$)          & 7,733       &  8,022         \\
			Number of users        & 269,849     & 528,719       \\
			Number of businesses   &111,267     &  108,936       \\ 
			Number of transactions & 12,993,179  &  24,507,586    \\
			\hline
		\end{tabular}
	\end{center}
\end{table}

The probability density function (PDF) of the number of transactions per user in 2011 and the amount of money spent per transaction is displayed in Figure \ref{Fig1}. We observe a strong heterogeneity among users regarding the number of transactions. The median value is 27 and the lower quartile is 8 in both provinces, the upper quartile is 69 for Barcelona and 66 for Madrid. Between 10 and 70 euros are spent in 50\% of the transactions with a median amount of 30 euros per transaction. 

For each transaction, the cost $d$ associated to a travel is estimated with the distance between the user's postcode of residence (lon/lat coordinates of the centroid) and the location of the business in which the transaction occurred (lon/lat of the credit card terminal). The value $v$ given to the travel purpose is inferred by the amount of money spent per transaction. The amount of money spent $v$ is divided into five intervals ($]0,50[$, $[50,75[$, $[75,100[$, $[100,200[$, $[200,500[$).

\begin{figure*}[!ht]
	\centering{\includegraphics[width=\linewidth]{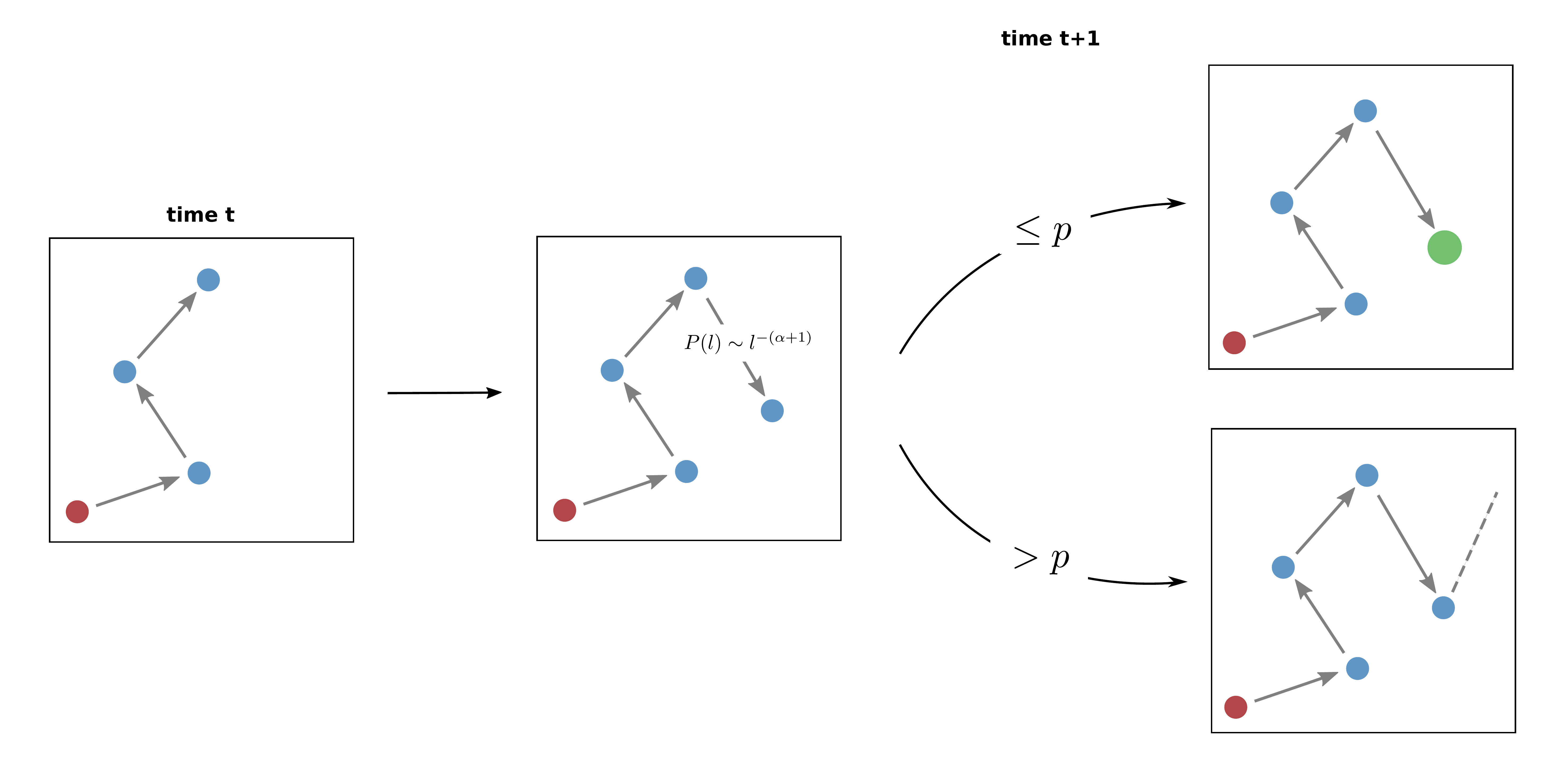}}
	\caption{\textbf{A schematic diagram of the model.} At each step, the individual leaves his/her actual location and moves in a random direction at a distance sample from a Pareto distribution $P(l)=\frac{\alpha {l_0}^\alpha}{l^{\alpha+1}}$. If the new location falls outside of the square boundaries the sampling process is repeated. According to the value $v$ given to the trip destination, the individual will then decide to stop or not his/her journey with a probability $p$. If the individual decides to end his/her journey the final destination is drawn at random in a circle of radius $r$ around the last position (green circle). \label{Fig2}}
\end{figure*}

\subsection*{Model}

The proposed model can be interpreted as a search process that stops when a satisfying object (destination) has been found \cite{Carra2016}. The rules of the model are outlined in Figure \ref{Fig2} . We assume that an individual starts the travel at his/her actual location (at home or work, for instance). The position of this first location is drawn at random in a square of lateral size $L$ expressed in kilometer. In practice, this parameter allows to take into consideration the geographical constraints of a case study site (administrative or geological boundaries for example). At each step, the individual move in a random direction and at a random distance sampled from a Pareto distribution $\displaystyle P(l)= \alpha \, {l_0}^\alpha \,  l^{- \alpha-1}$, where $\alpha$ is the exponent and $l_0$ the minimum spatial scale considered. At each step, the possibility to end the travel is represented by a probability $p$ of fulfilling the trip goal. Note that unlike most of the models described in the introduction, since only short-range mobility patterns are considered, our model does not take explicitly into account time. The probability of stopping $p$ is assumed to have an inverse relation with the importance given to the trip goal $v$. The higher the value $v$ associated to the objective of the travel, the longer the search process (i.e., low value of $p$) and the higher the distance $d$ between origin and destination  can become. If the purpose of the trip is a search to buy an object, the individuals would be willing to explore more shops or to travel further as the item price increases (buying a car requires more  \enquote{energy} than a piece of bread). Finally, when the individual decides to end his/her journey the final destination is drawn at random in a circle of radius $r$ around the last position. This mechanism is included to take into account the uncertainty present in the data on the exact position of the retailing center. Still, in case of a more abstract framework the model could be simplified by making $r \to 0$ and setting the purchase place in the current agent's location.

\subsection*{Model calibration}

The comparison between model and data is based on the probability density function (PDF) of the distance $d$ between the origin and the destination. The simulated PDF is obtained with 100,000 simulations of the model. The model has five parameters $L$, $\alpha$, $l_0$, $r$ and $p$. $L$ controls the size of the modeling area, $l_0$ and $r$ the minimum spatial resolution,  $\alpha$ the jump length and $p$ the users' tendency to explore the modeling area. The free parameter $l_0$ and $r$ allows notably to control the users' exploration behavior at short distance from home. The numerical values of these five parameters were determined from the empirical data in two steps and independently for both provinces. We first adjusted the five parameters' value by minimizing the Kolmogorov-Smirnov distance between observed and simulated PDF of $d$ (all values of $v$ combined). We then adjusted the value of $p$ according to the PDF of $d$ for each interval of amount of money spend $v$.

\begin{figure*}[!ht]
	\centering{\includegraphics[width=\linewidth]{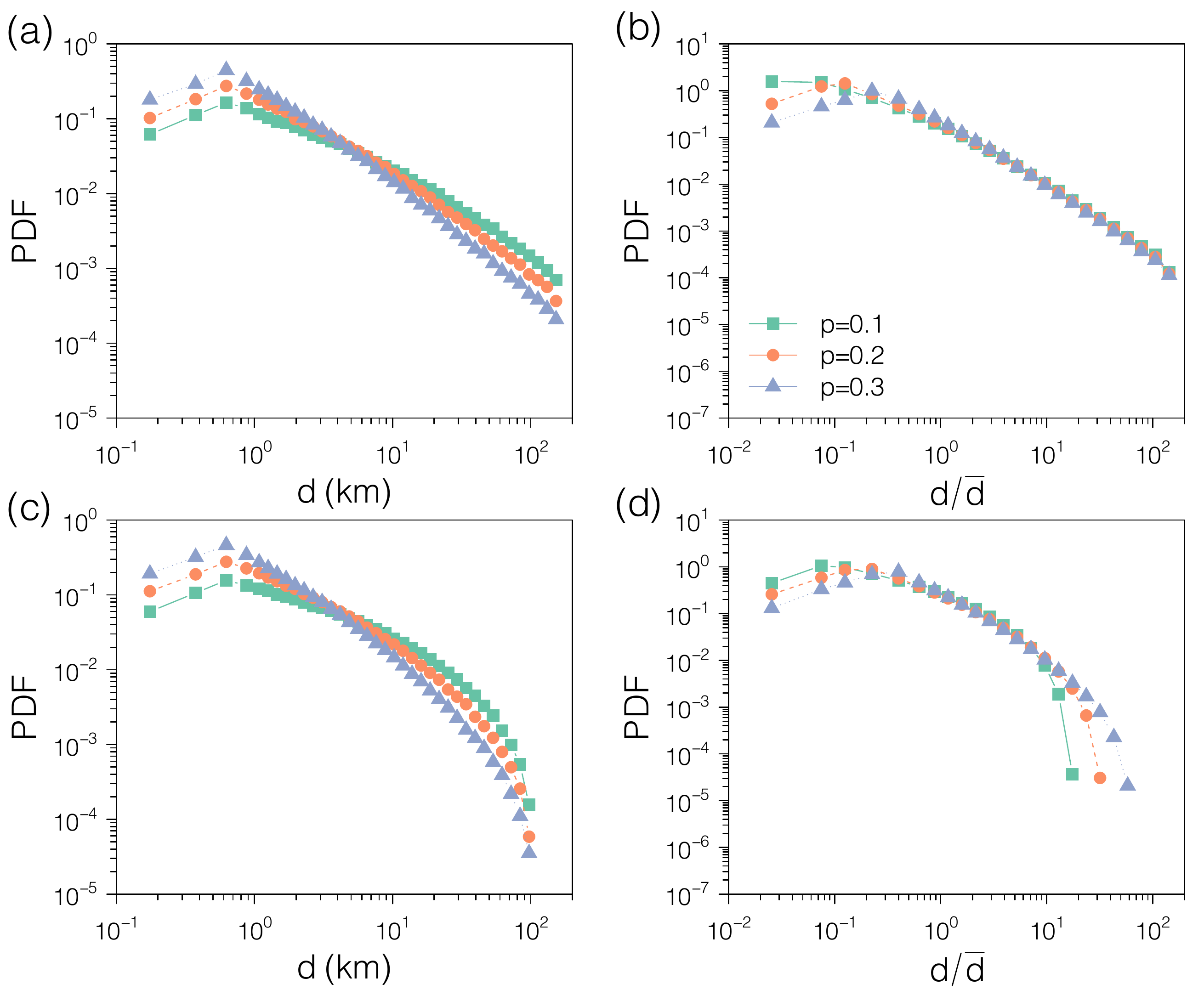}}
	\caption{\textbf{Probability density function of the distance $d$ from the residence and the final location as a function of $p$ obtained with the calibrated model for Barcelona.} The original distributions are shown in (a) and (c), while the normalized distributions obtained by dividing by the median distance $\bar{d}$ are in (b) and (e).  In (a) and (b), there is no bounding box, $L=\infty$, while in (c) and (d) $L=100$ km. \label{Fig3}}
\end{figure*}

\subsection*{Model features}

Every agent in the model is performing a short L\'evy flight in the limited space provided by the box of side $L$. To better understand the model features, it is helpful to to take first the limits $L\to \infty$, so there is no spatial constraint. The L\'evy flight is not, actually, complete because we introduce a stopping mechanism with $p$. It implies that the number of jumps, $n$, that an agent takes follows the geometric distribution:
\begin{equation}
P_n(n) = (1-p)^{n-1} \, p .
\end{equation}
The average number of jumps is thus given by $\langle n \rangle = 1/p$. Lower $p$ means more jumps and, therefore, the potential for longer distances in the distribution of distance from the origin to the purchase location, $P(d)$. Recalling that $p$ will be related to $v$ with an inverse function, larger $v$ implies as well longer distances $d$. The distribution $P(d)$ comes thus from the aggregation of a finite number $n$ of L\'evy jumps. In the limit $n \to \infty$, it could be expressed in function of the L\'evy $\alpha$-stable distributions. On the contrary, for small $n$ the analytical expression of $P(d)$ does not correspond to L\'evy's generalization of the central limit theorem. In any case, there is an inverse relation between the median of the distance $d$, $\bar{d}$, and $p$. Examples of the distributions $P(d)$ for different values of $p$ can be seen in Figure \ref{Fig3}(a). The range of small $d$ values is flattened by the presence of a minimal scale (controlled by $l_0$ and  $r$), whereas as expected a long power-law like tail appears for large $d$. In this case, there is no other characteristic scales in the model beyond the small scale and $\bar{d}$ as shown by the collapse of the curves for different $p$ obtained dividing the x-axis by $\bar{d}$ and normalizing the distributions again (Figure \ref{Fig3}(b)).

\begin{figure*}[!ht]
	\centering{\includegraphics[width=\linewidth]{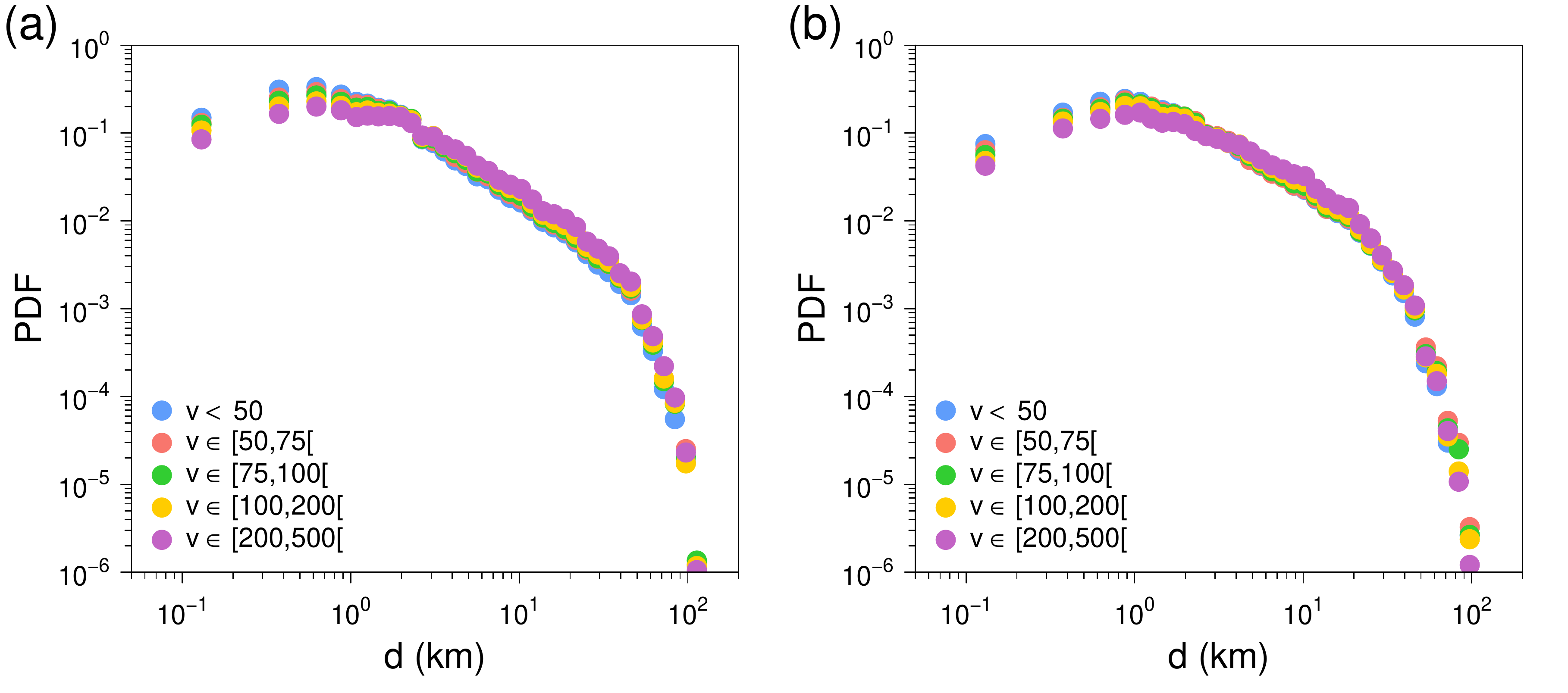}}
	\caption{\textbf{Distribution of the distance traveled according to the amount of money spent in Barcelona (a) and Madrid (b).} Probability density function of the distance between the users' place of residence and the location of the business in which the transaction occurred according to the amount of money spent $v$. The color of the curves represents different ranges of amount of money spent. \label{Fig4}}
\end{figure*}

The picture changes if $L$ and, consequently, the bounding box is finite. The limited L\'evy flights are then occurring inside a constrained space and those jumping outside are not considered. This introduces a maximum scale in the $P(d)$ distributions as can be seen in Figure \ref{Fig3}(c), which manifests in a fast (exponential-like) decay for large values of $d$. The curves still maintained the power law-like properties and the possibility of collapse dividing $d$ by $\bar{d}$, but in a restricted domain of $d$ values (see Figure \ref{Fig3}(d)).

\section*{Results}

We start by exploring empirically the relationship between the travel distance $d$ and the importance given to its destination $v$. Figure \ref{Fig4} displays the probability density function of the distance between the users' home and the location of the business in which the transaction occurred according to the amount of money spent $v$ divided into five intervals. Several regimes can be observed. First, the probability to travel a certain distance to make a purchase increases, reaching a maximum between $500$ m and $1$ km, and, then, the probability starts to decrease, slowly at first, and then more rapidly, exhibiting a power-law like decay. Finally, after $20-50$ km the province boundaries act as a natural cutoff in the distribution (our data is limited to single provinces). The shape of distribution is very similar for each range of amount of money spent. It seems however that the distance traveled globally increases with the amount of money spent. 

\begin{figure*}[!ht]
	\includegraphics[width=\linewidth]{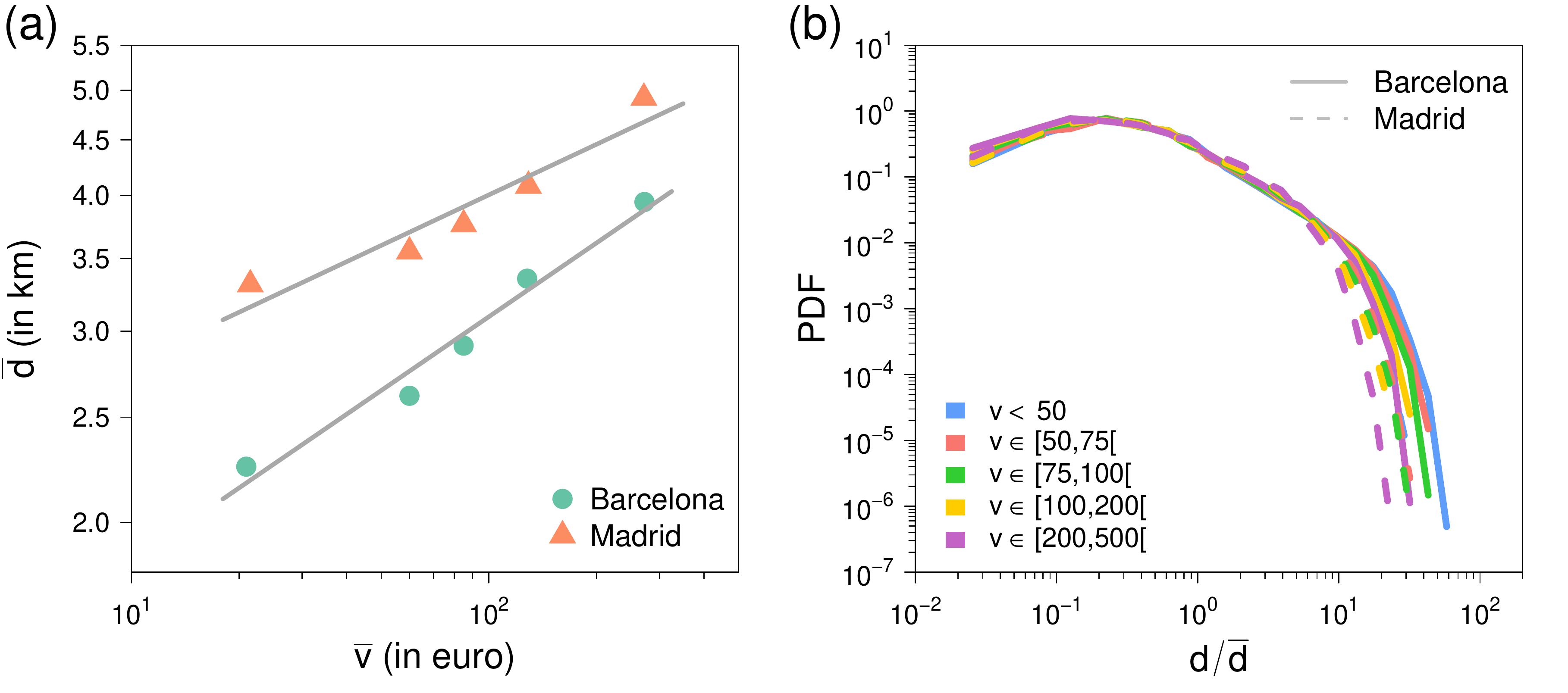}
	\caption{\textbf{Scaling relationship between amount of money spent and distance traveled.} (a) Median distance $\bar{d}$ as a function of the median amount of money spent $\bar{v}$ for each range of prices in Barcelona (green dots) and Madrid (orange triangles). (b) Probability density function of the distance normalized by the median distance according to the amount of money spent $v$ in Barcelona (solid line) and Madrid (dashed line). The color of the curves represents different ranges of amount of money spent. \label{Fig5}}
\end{figure*}

Figure \ref{Fig5}a shows for each range of values $v$ the median distance traveled $\bar{d}$ as a function of the median amount of money spent $\bar{v}$. Although the distance traveled is globally higher in Madrid than in Barcelona, the distance traveled increases with the amount of money spent following the scaling relationship $\bar{d} \sim \bar{v}^{\gamma}$ in both provinces. We obtain a value $\gamma=0.23 \pm 0.02$ for Barcelona and $\gamma=0.15 \pm 0.03$ for Madrid estimated with a log-log regression. It is interesting to note that this relation between $d$ and $v$ seems to be the unique driver of the differences observed between the PDFs in Figure \ref{Fig4}. Reminding the collapse in the model in Figure \ref{Fig3}, as can be seen in Figure \ref{Fig5}b a scaling factor depending only on $\bar{d}$ can be used to visually collapse all the PDFs shown in Figure \ref{Fig5}b into a single curve (except for the maximum values constrained by the provinces' geographical boundaries). This suggests that the mechanisms underlying trip generation are the same for all price ranges and the only difference is a characteristic distance $\bar{d}$, which is a function of the price $v$ of the item to purchase. 

\begin{figure}[!h]
	\centering
	\includegraphics[width=\linewidth]{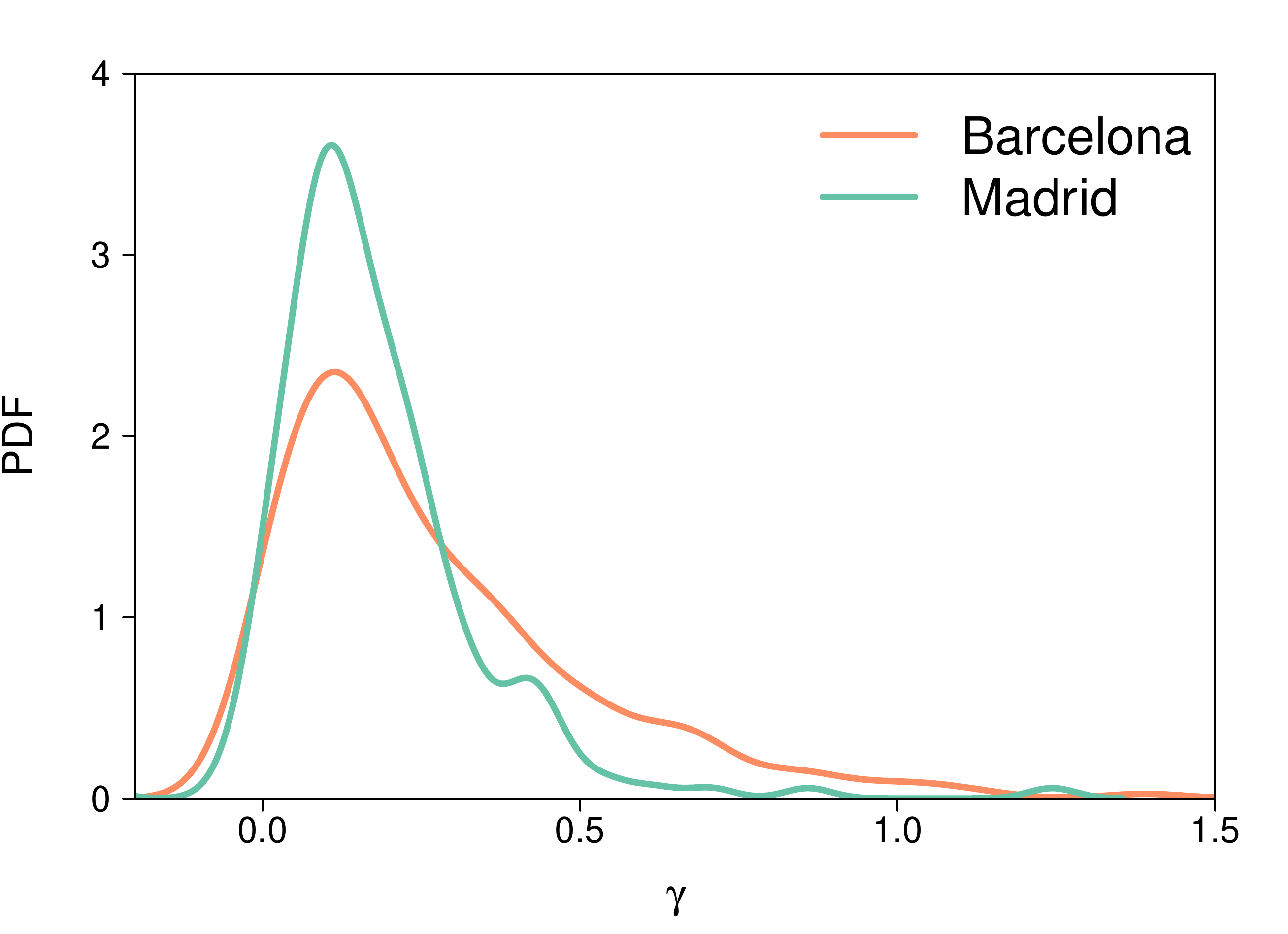}
	\caption{\textbf{Probability density function of the exponent $\pmb{\gamma}$ estimated with a log-log model for each postcode in the provinces of Barcelona and Madrid.} \label{Fig6}}
\end{figure}

Nevertheless, we need to verify that this result holds whenever the spatial distributions and types of users and businesses. To ensure that it is the case, we plot in Figure \ref{Fig6} the distribution of the exponent $\gamma$ estimated with a log-log model for each postcode in both provinces. The value of $\gamma$ is globally strictly higher than 0, suggesting that the positive correlation between $\bar{d}$ and $\bar{v}$ does not dependent of the user's postcode of residence. Moreover, this positive correlation between the two quantities is independent of the users' sociodemographic characteristics (gender, age and occupation) and the business categories (see the SI for more details).

\begin{figure*}[!ht]
	\includegraphics[width=\linewidth]{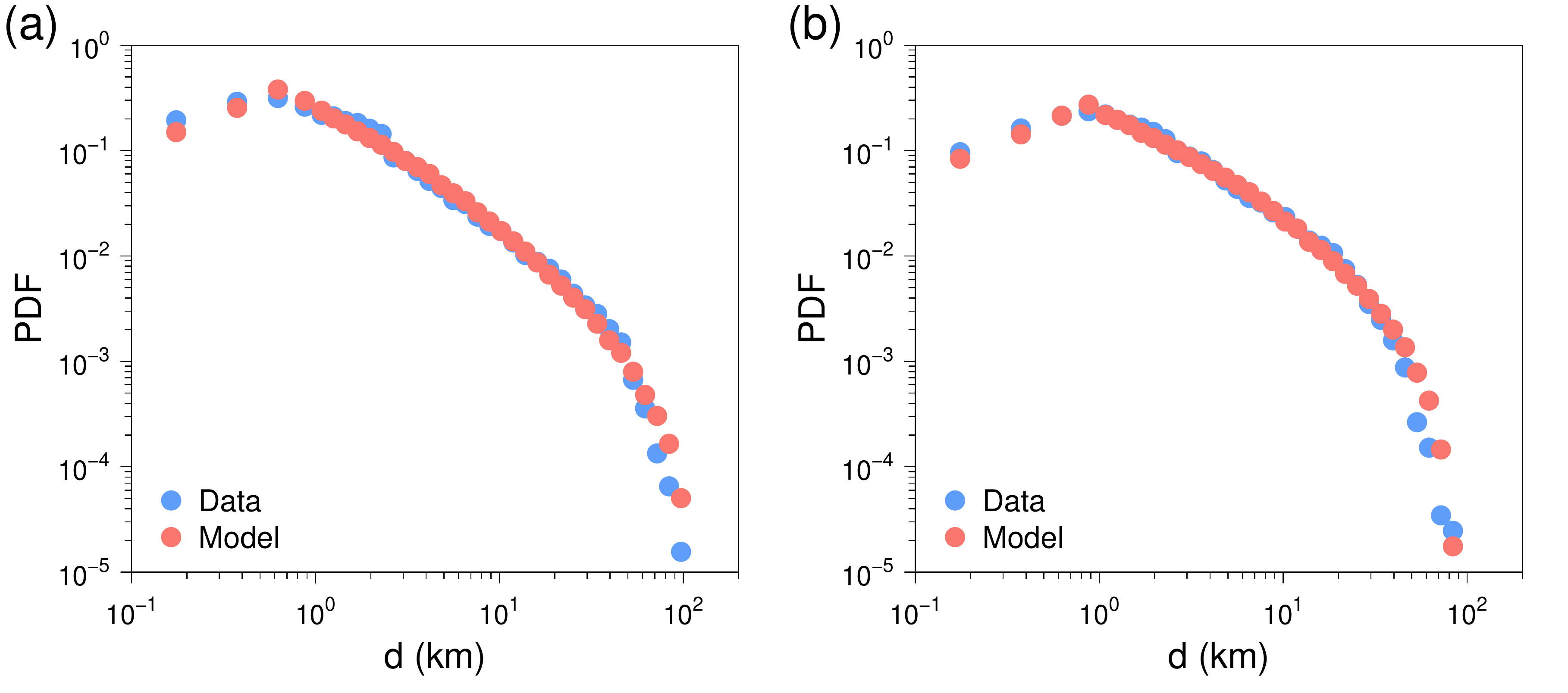}
	\caption{\textbf{Comparison between data and model.} Probability density function of the distance between the users' place of residence and the location of the business in which the transaction occurred (all of the amounts combined) obtained with the data (in blue) and the calibrated model (in red). (a) In Barcelona the best results are obtained with a square of side length $L=100$ km, a radius $r$ equal to $300$ meters, mobility parameters $l_0=300$ meters, $\alpha=0.6$ and $p=0.3$. (b) In Madrid the best results are obtained with a square of side length $L=75$ km, a radius $r$ equal to $400$ meters, mobility parameters $l_0=400$ meters, $\alpha=0.6$ and $p=0.25$. \label{Fig7}}
\end{figure*}

We now focus on the results obtained with our model in order to reproduce and explain the relationship between $d$ and $v$ observed in the data. As described in Material and Methods, we first consider the distribution of all the amounts combined in order to calibrate the five parameters. As it can be seen in Figure \ref{Fig7}, the fit is quite good. We obtain similar results in both provinces. The modeling area represented by a square of lateral size $L$ is bigger in Barcelona (100 km) than in Madrid (75 km). The parameter $\alpha$, exponent of the Pareto distribution, is equal to $0.6$ which is consistent with values obtained in other studies \cite{Brockmann2006,Gonzalez2008,Hawelka2014,Lenormand2015b,Cheng2011,Raichlen2014}. We obtain a value of $p$ equal to $0.3$ in Barcelona and $0.25$ in Madrid, this value, comprised between $0$ and $1$, has an inverse relation to the energy that people are willing to invest in order to go shopping in both provinces. 

\begin{figure*}[!ht]
	\includegraphics[width=\linewidth]{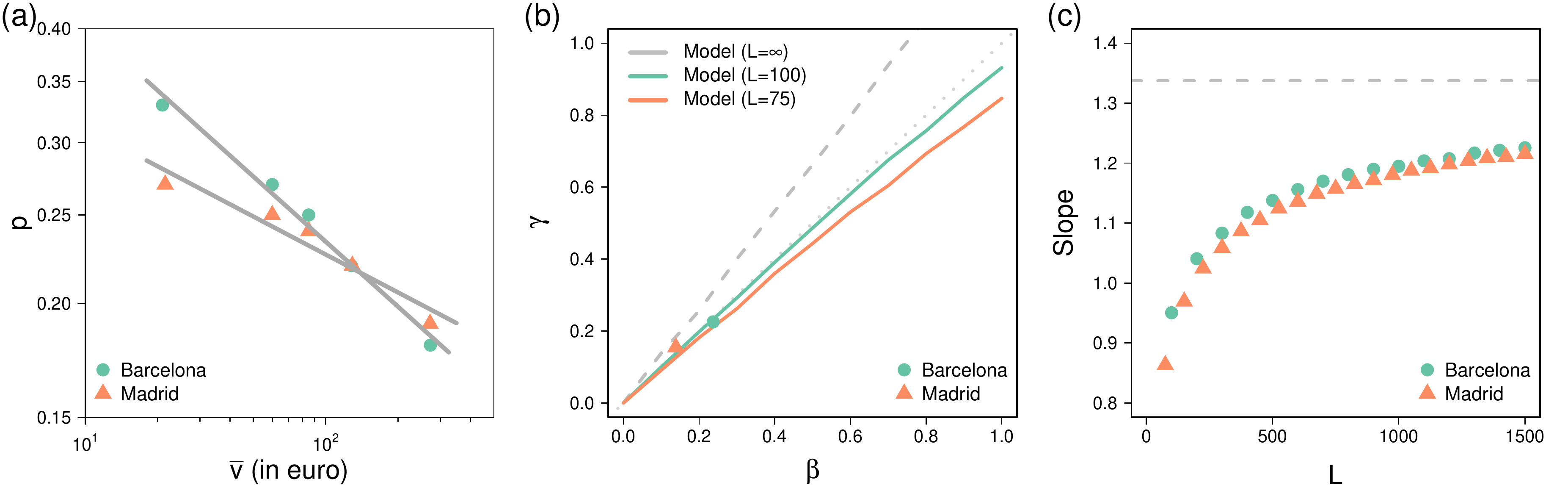}
	\caption{\textbf{Relationship between $\pmb{\gamma}$ and $\pmb{\beta}$.} (a) Probability $p$ as a function of the amount of money spent $v$ in each bin in Barcelona (green dots) and Madrid (orange triangles). (b) Relation between $\gamma$ and $\beta$ obtained with the model and the data for Barcelona (in green) and Madrid (in orange). The lines correspond to the simulations and the symbol to the data (Barcelona in green and Madrid in orange). The grey solid line corresponds to the results obtained with the model with $L=\infty$. (c) Evolution of the slope between $\gamma$ and $\beta$ as a function of $L$ for Barcelona (green dots) and Madrid (orange triangles). \label{Fig8}}
\end{figure*}

Finally, we explore the behavior of $p$ according to the median amount of money spent $\bar{v}$. The results obtained are plotted in Figure \ref{Fig8}a. As expected, the value of $p$ decreases with increasing $v$, which implies that the distance traveled grows with the price of the item to purchase. Furthermore, we find that a scaling relation of the type $p \sim \bar{v}^{-\beta}$ adjusts well to the data. We obtain a value $\beta=0.24 \pm 0.01$ for Barcelona and $\beta=0.14 \pm 0.02$ for Madrid estimated with a log-log regression. However, keep in mind that the model does not impose a given relation between $p$ and $\bar{v}$, it can be general with different type of data leading to diverse relationships (or exponents if the power-law scaling holds). In our case, both $\bar{d}$ and $p$ can be expressed as scaling functions of $\bar{v}$. It is, therefore, important to understand the relation between the direct observable in mobility $\bar{d}$ and our model's $p$. If the basic displacement distribution had had a finite second moment, i.e., the movement was a random walk in 2D, it would have been possible to find analytical approximations for the final distance. However, this task becomes complex with a finite number of steps in a L\'evy flight. 

Therefore, to analyze the relationship between $\gamma$ and $\beta$, we assume a relation $p \sim \bar{v}^{-\beta}$ to generate five $p$ values for a given $\beta$ value. We normalize the five $p$ values in order to preserve the average value observed in the data (i.e. values displayed in Figure \ref{Fig8}a). We then simulate the five $\bar{d}$ value associated with the $p$ values with our model (using the calibrated $L$, $\alpha$, $l_0$ and $r$ values for both provinces). We finally estimate the exponent $\gamma$ from $\bar{d} \sim \bar{v}^{\gamma}$ and compare the values of $\gamma$ obtained versus those of the corresponding $\beta$. The results of this exercise are shown in Figure \ref{Fig8}b. The relation is linear and close to the identity ($\sim 0.9$) but we note a slight difference between the results obtained with the model parameterization used for Barcelona and Madrid. This is mainly due to the size of modeling area $L$. The grey dashed line in Figure \ref{Fig8}b represents the relationship between $\gamma$ and $\beta$ obtained with an infinite modeling area. The effect of increasing the value of $L$ on the slope of the linear relationship is also exposed in Figure \ref{Fig8}c. We observe that the slope ranges from $0.85$ to $1.1$ for value of $L$ lower than 500 km, after that it increases slowly until reaching the asymptotic value $1.33$.  
This change in the slope is essentially due to the progressive reduction of the sum of L\'evy flights' truncation. It is worth noting that the slope obtained with values of $L$ reflecting an intra-urban mobility scale are very closed to one. An equality between $\gamma$ and $\beta$ is also consistent with the empirical observations made in Barcelona and Madrid (Figure \ref{Fig8}b), suggesting that the probability to stop the journey could be inversely proportional to the median distance traveled ($p \sim 1/\bar{d}$).

\section*{Discussion}

In summary, we introduced a model of individual human mobility patterns able to reproduce and explain the relationship between the travel cost associated to a trip and the importance given to its destination observed in credit card data recorded in the provinces of Barcelona and Madrid in 2011. In particular, we have shown that the distances between place of residence and place of purchase increase with the amount of money spent following a similar scaling relationship in both provinces. The model that we proposed is able to reproduce these behaviors and also to mimic the final scaling relation.   

Overall, we observed a good agreement between the results obtained in Barcelona and Madrid. Both provinces show similar trends in the relationship between amount of money spent and distance traveled. The exponent values observed in the scaling relationships are of the same order of magnitude in both provinces and the calibrated parameter values obtained with the model are also very similar. More research is needed to elucidate whether the patterns found are common to other countries and cities, and, specially, whether the small differences are related to the diverse city shapes or the geographical structure of the administrative units.

\subsection*{Limitations of the study}

The results obtained give a good confidence in the robustness of the scaling relationship observed in the data by assessing the effect of the users' characteristics and business category on the exponent of the scaling relationship in the two provinces. However, it will be important to evaluate our hypothesis and our model on case studies coming from other countries/continents and on different data sources. A limitation of the study lies in the nature of our data samples spatially constrained by the province boundaries. This forces us to include in the model a bounding box and restricts our capability to reach analytical results. As in \cite{Louail2017}, we also assumed that every shopping trip starts at home and ends at the place of purchase without considering more complicated case including sequences of purchases. Nevertheless, we replicated the analysis considering, for each user, only days with a unique transaction and we obtained very similar results (see Figure S3 in SI). Finally, online shopping could be also a handicap for our analysis. Unfortunately, we cannot distinguish online and offline shops in our data. Online shopping has become more relevant with time, its presence was not so strong in the shopping mixing of 2011 as today, and the retailer must be included in the same province is a strict limitation for most of the online shops. In the online purchases, the distance should not be anymore an important variable and it should not show a clear relation with v, given that it is possible to buy equally items of any price. 

\subsection*{Concluding Remarks}

To conclude, this study is a first attempt to quantify the relationship between travel cost associated to a trip and importance given to its destination. The results obtained in this study shed new light on the modeling of human mobility patterns at an individual scale. We are quite aware that trip motivations are very complicated to quantify but we truly believe that it is an important topic. An accurate modeling of daily human mobility patterns in cites is crucial in a wide range of applications. Getting better insights on the relationship between trip characteristics and travel motivations would allow to gain a better understanding of urban dynamics in order to optimize cities. We hope that more and more (hopefully open) data will be made available in years to come to study the importance of trip destination and its role in the modeling of human mobility patterns.

\section*{Acknowledgements} 

ML thanks the French National Research Agency for its financial support (project NetCost, ANR-17-CE03-0003 grant). MSM and JJR acknowledge partial funding from the Spanish Ministry of Science, Innovation and Universities, the National Agency for Research Funding AEI and FEDER (EU) under the grant PACSS (RTI2018-093732-B-C22) and the Maria de Maeztu program for Units of Excellence in R\&D (MDM-2017-0711). We also thank Teodoro Dannemann for fruitful discussions.

\bibliographystyle{unsrt}
\bibliography{DistAmount}

\begin{thebibliography}{10}

\bibitem{Barbosa2018}
H.~Barbosa, M.~Barthelemy, G.~Ghoshal, C.~R. James, M.~Lenormand, T.~Louail,
  R.~Menezes, J.~J. Ramasco, F.~Simini, and M.~Tomasini.
\newblock Human mobility: Models and applications.
\newblock {\em Physics Reports}, 734:1--74, 2018.

\bibitem{Brockmann2006}
D.~Brockmann, L.~Hufnagel, and T.~Geisel.
\newblock The scaling laws of human travel.
\newblock {\em Nature}, 439(7075):462--465, 2006.

\bibitem{Gonzalez2008}
M.~C. Gonz{\'a}lez, C.~A. Hidalgo, and A.-L. Barab\'{a}si.
\newblock {Understanding individual human mobility patterns}.
\newblock {\em Nature}, 453(7196):779--782, 2008.

\bibitem{Hawelka2014}
B.~Hawelka, I.~Sitko, E.~Beinat, S.~Sobolevsky, P.~Kazakopoulos, and C.~Ratti.
\newblock Geo-located twitter as a proxy for global mobility patterns.
\newblock {\em Cartography and Geographic Information Science}, 41:260--271,
  2014.

\bibitem{Lenormand2015b}
M.~Lenormand, B.~Goncalves, A.~Tugores, and J.~J. Ramasco.
\newblock Human diffusion and city influence.
\newblock {\em Journal of Royal Society Interface}, 12:20150473, 2015.

\bibitem{Cheng2011}
Z.~Cheng, J.~Caverlee, K.~Lee, and D.~Sui.
\newblock Exploring millions of footprints in location sharing services.
\newblock {\em Proceedings of the Fifth International AAAI Conference on
  Weblogs and Social Media}, 2011.

\bibitem{Raichlen2014}
D.~A. Raichlen, B.~M. Wood, A.~D. Gordon, A.~Z.~P. Mabulla, F.~W. Marlowe, and
  H.~Pontzer.
\newblock Evidence of {L{\'e}vy} walk foraging patterns in human
  hunter-gatherers.
\newblock {\em Proceedings of the National Academy of Sciences},
  111(2):728--733, 2014.

\bibitem{Roth2011}
C.~Roth, S.~M. Kang, M.~Batty, and M.~Barthelemy.
\newblock {Structure of Urban Movements: Polycentric Activity and Entangled
  Hierarchical Flows}.
\newblock {\em PLoS ONE}, 6(1):e15923, 2011.

\bibitem{Hasan2013}
S.~Hasan, C.~M. Schneider, S.~V. Ukkusuri, and M.~C. Gonz{\'a}lez.
\newblock Spatiotemporal patterns of urban human mobility.
\newblock {\em Journal of Statistical Physics}, 151(1-2):304--318, 2013.

\bibitem{Song2010a}
C.~Song, Z.~Qu, N.~Blumm, and A.-L. Barab\'{a}si.
\newblock {Limits of Predictability in Human Mobility}.
\newblock {\em Science}, 327(5968):1018--1021, 2010.

\bibitem{Schneider2013}
C.~M. Schneider, V.~Belik, T.~Couronn{\'{e}}, Z.~Smoreda, and M.~C.
  Gonz{\'{a}}lez.
\newblock {Unravelling daily human mobility motifs}.
\newblock {\em Journal of The Royal Society Interface}, 10(84):20130246, 2013.

\bibitem{Kang2012}
C.~Kang, X.~Ma, D.~Tong, and Y.~Liu.
\newblock {Intra-urban human mobility patterns: An urban morphology
  perspective}.
\newblock {\em Physica A: Statistical Mechanics and its Applications},
  391(4):1702--1717, 2012.

\bibitem{Lotero2014}
L.~Lotero, A.~Cardillo, R.~Hurtado, and J.~Gomez-Gardenes.
\newblock Several multiplexes in the same city: The role of socioeconomic
  differences in urban mobility.
\newblock {\em Available at SSRN 2507816}, 2014.

\bibitem{Lenormand2015a}
M.~Lenormand, T.~Louail, O.~Garcia~Cant{\'u}, M.~Picornell, R.~Herranz,
  M.~Barthelemy, M.~San~Miguel, and J.~J. Ramasco.
\newblock Influence of sociodemographic characteristics on human mobility.
\newblock {\em Scientific Reports}, 5:10075, 2015.

\bibitem{Gauvin2019}
L.~Gauvin, M.~Tizzoni, S.~Piaggesi, A.~Young, N.~Adler, S.~Verhulst, L.~Ferres,
  and C.~Cattuto.
\newblock Gender gaps in urban mobility.
\newblock \url{https://arxiv.org/abs/1906.09092}, 2019.

\bibitem{Wang2011}
D.~Wang, D.~Pedreschi, C.~Song, F.~Giannotti, and A.-L. Barab\'{a}si.
\newblock Human mobility, social ties, and link prediction.
\newblock In {\em Proceedings of the 17th ACM SIGKDD International Conference
  on Knowledge Discovery and Data Mining}, KDD '11, pages 1100--1108, 2011.

\bibitem{Grabowicz2014}
P.~A Grabowicz, J.~J. Ramasco, B.~Gon{\c{c}}alves, and V.~M Egu{\'\i}luz.
\newblock Entangling mobility and interactions in social media.
\newblock {\em PLoS ONE}, 9:e92196, 2014.

\bibitem{Picornell2015}
M.~Picornell, T.~Ruiz, M.~Lenormand, J.~J. Ramasco, T.~Dubernet, and
  E.~Fr{\'i}as-Mart{\'i}nez.
\newblock Exploring the potential of phone call data to characterize the
  relationship between social network and travel behavior.
\newblock {\em Transportation}, 42(4):647--668, 2015.

\bibitem{Toole2015}
J.~L. Toole, C.~Herrera-Yaq{\"u}e, C.~M. Schneider, and M.~C. Gonz{\'a}lez.
\newblock Coupling human mobility and social ties.
\newblock {\em Journal of The Royal Society Interface}, 12(105):20141128, 2015.

\bibitem{Rhee2008}
I.~Rhee, M.~Shin, S.~Hong, K.~Lee, and S.~Chong.
\newblock On the levy-walk nature of human mobility.
\newblock In {\em INFOCOM 2008. The 27th Conference on Computer Communications.
  IEEE}, 2008.

\bibitem{Song2010b}
C.~Song, T.~Koren, P.~Wang, and A.-L. Barab\'{a}si.
\newblock {Modelling the scaling properties of human mobility}.
\newblock {\em Nature Physics}, 6(10):818--823, 2010.

\bibitem{Lee2009}
K.~Lee, S.~Hong, S.~J. Kim, I.~Rhee, and S.~Chong.
\newblock {SLAW: A Mobility Model for Human Walks}.
\newblock In {\em Proceedings of the 28th Annual Joint Conference of the IEEE
  Computer and Communications Societies (INFOCOM)}, 2009.

\bibitem{wolf2001}
J.~Wolf, R.~Guensler, and W.~Bachman.
\newblock Elimination of the travel diary: Experiment to derive trip purpose
  from global positioning system travel data.
\newblock {\em Transportation Research Record}, 1768:125--134, 2001.

\bibitem{bohte2009}
W.~Bohte and K.~Maat.
\newblock Deriving and validating trip purposes and travel modes for multi-day
  gps-based travel surveys: A large-scale application in the netherlands.
\newblock {\em Transportation Research Part C: Emerging Technologies},
  17:285--297, 2009.

\bibitem{Parsons2003}
G.~R. Parsons.
\newblock The {Travel} {Cost} {Model}.
\newblock In P.~A. Champ, K.~J. Boyle, and T.s~C. Brown, editors, {\em A
  {Primer} on {Nonmarket} {Valuation}}, The {Economics} of {Non}-{Market}
  {Goods} and {Resources}. Springer Netherlands, Dordrecht, 2003.

\bibitem{Butterfield2016}
B.~J. Butterfield, A.~L. Camhi, R.~L. Rubin, and C.~R. Schwalm.
\newblock Chapter {Five} - {Tradeoffs} and {Compatibilities} {Among}
  {Ecosystem} {Services}: {Biological}, {Physical} and {Economic} {Drivers} of
  {Multifunctionality}.
\newblock In G.~Woodward and D.~A. Bohan, editors, {\em Advances in
  {Ecological} {Research}}, volume~54 of {\em Ecosystem {Services}: {From}
  {Biodiversity} to {Society}, {Part} 2}, pages 207--243. Academic Press, 2016.

\bibitem{Viswanathan1996}
G.~M. Viswanathan, V.~Afanasyev, S.~V. Buldyrev, E.~J. Murphy, P.~A. Prince,
  and H.~E. Stanley.
\newblock {L\'{e}vy flight search patterns of wandering albatrosses}.
\newblock {\em Nature}, 381(6581):413--415, 1996.

\bibitem{Viswanathan1999}
G.~M. Viswanathan, S.~V. Buldyrev, S.~Havlin, M.~G.~E. da~Luz, E.~P. Raposo,
  and H.~E. Stanley.
\newblock Optimizing the success of random searches.
\newblock {\em Nature}, 411:911--914, 1999.

\bibitem{Viswanathan2010}
G.~M. Viswanathan.
\newblock Ecology: {Fish} in {L{\'e}vy}-flight foraging.
\newblock {\em Nature}, 465(7301):1018--1019, 2010.

\bibitem{Louail2017}
T.~Louail, M.~Lenormand, J.~M. Arias, and J.~J. Ramasco.
\newblock Crowdsourcing the {{Robin Hood}} effect in cities.
\newblock {\em Applied Network Science}, 2(1):11, 2017.

\bibitem{Carra2016}
G.~Carra, I.~Mulalic, M.~Fosgerau, and M.~Barthelemy.
\newblock Modelling the relation between income and commuting distance.
\newblock {\em Journal of The Royal Society Interface}, 13(119):20160306, 2016.

\end{thebibliography}

\onecolumngrid
\vspace*{2cm}
\newpage

\makeatletter
\renewcommand{\fnum@figure}{\sf\textbf{\figurename~\textbf{S}\textbf{\thefigure}}}
\renewcommand{\fnum@table}{\sf\textbf{\tablename~\textbf{S}\textbf{\thetable}}}
\makeatother

\setcounter{figure}{0}
\setcounter{table}{0}
\setcounter{equation}{0}

\section*{Appendix}

\subsection*{Data preprocessing}

As mentioned in the main text, we analyzed in this study a credit card dataset containing information about $35$ million bank card transactions made by credit card users of the Banco Bilbao Vizcaya Argentaria (BBVA) in the province of Barcelona and Madrid in $2011$. The dataset used in this study has been already presented and analyzed in \citep{Lenormand2015a}. We only applied two filters, one on the average number of transactions per day and another one on the maximum amount of money spent per transaction. 

First, we filtered out users with a number of transactions per day higher than three. We determined this threshold by plotting the number of transactions per user as a function of the number of days with at least one transaction. We observe in Figure S1 that most of the users made less than three transactions per day (red line). Only a few users (234 for Barcelona and 613 for Madrid) made more than three transactions per day in 2011 representing less than 0.12\% of the users in both case studies. 

\begin{figure}[!h]
	\centering
	\includegraphics[width=14cm]{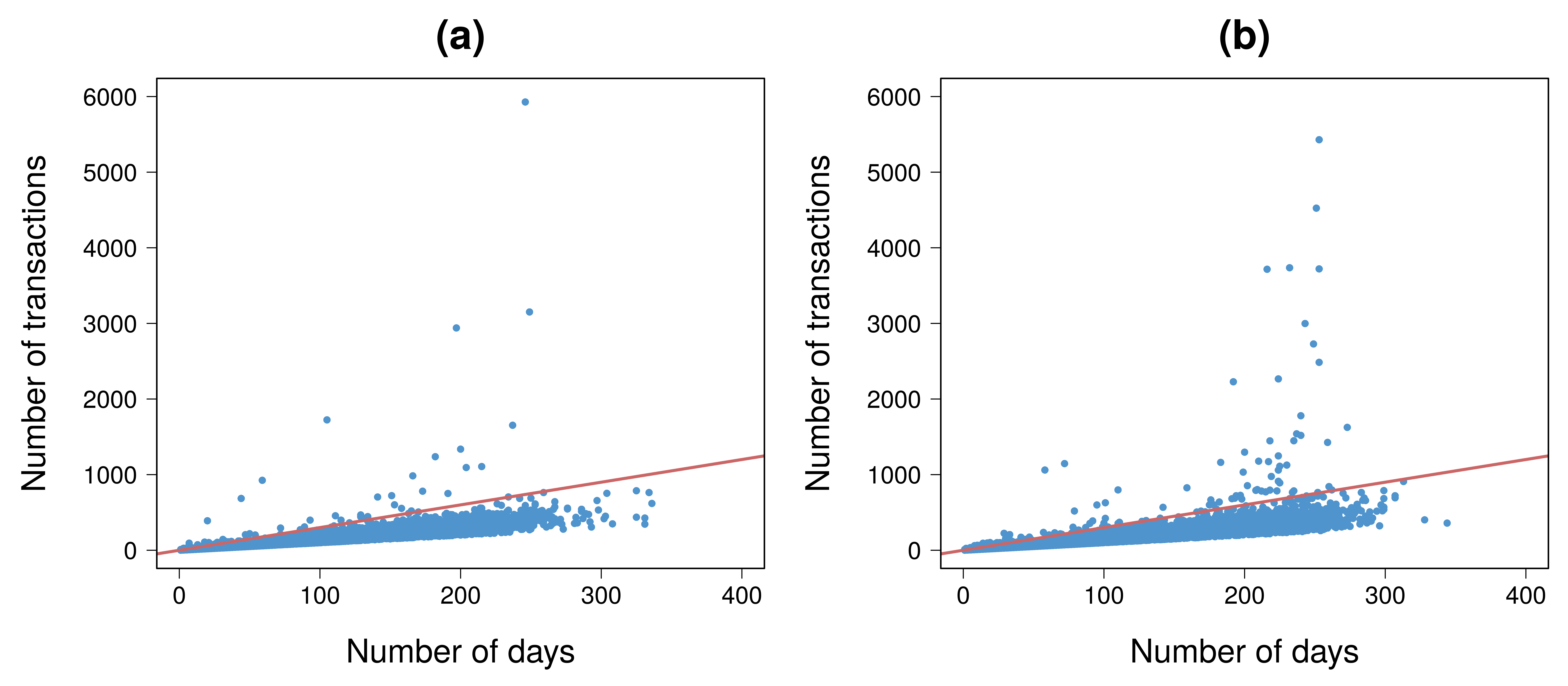}
	\caption{\textbf{Number of transactions as a function of the number of days with at least one transaction in 2011 in Barcelona (a) and Madrid (b).} Each blue dot represents a user. The red line represents the threshold of three transactions per day. \label{FigS1}}
\end{figure}

Then we removed from the database all the transaction  with an amount higher than $500$ euros (85,665 for Barcelona and 183,155 for Madrid) representing less than 0.75\% of the transactions in both case studies.

\subsection*{Effect of the users' characteristics on the exponent $\gamma$}

As mentioned in the main text the relationship between the median distance $\bar{d}$ and the median amount of money spent $\bar{v}$ can be well-approximated by a log-log function. However, we need to verify that this positive correlation between the two quantities does not depend of the users' sociodemographic characteristics. Each BBVA user is connected with sociodemographic characteristics (gender, age and occupation). For the sake of convenience, we consider five age groups ($]15,30]$, $]30,45]$, $]45,60]$, $]60,75]$, $>75$) and five types of occupations (student, unemployed, employed, homemaker, and retired). The relationship between the distance traveled $\bar{d}$ and the amount of money spent $v$ according to the users' sociodemographic characteristics is displayed in Table S1. Here again, the value of $\gamma$ is always strictly higher than 0.

\begin{table}[!h]
	\caption{\textbf{Relationship between the distance traveled and the amount of money spent according to the users' economic and sociodemographic characteristics.}}
	\label{TabS1}
	\begin{center}
		\begin{tabular}{lccc}
			\hline
			\centering Category    & Median Distance (BCN / MAD)  & Slope (BCN / MAD) & R$^2$ (BCN / MAD)  \\
			\hline
			
			Total  &  2.41 / 3.49  &  0.23 / 0.15  &  0.97 / 0.91 \\
			Men  &  2.98 / 3.97  &  0.2 / 0.15  &  0.99 / 0.87 \\
			Women  &  2.11 / 3.11  &  0.2 / 0.13  &  0.93 / 0.84 \\
			Age $\in$ ]15,30]  &  3.14 / 4.65  &  0.17 / 0.13  &  0.99 / 0.91 \\
			Age $\in$ ]30,45]  &  2.52 / 3.92  &  0.26 / 0.18  &  0.98 / 0.86 \\
			Age $\in$ ]45,60]  &  2.15 / 2.75  &  0.26 / 0.24  &  0.96 / 0.96 \\
			Age $\in$ ]60,75]  &  1.84 / 2.21  &  0.28 / 0.27  &  0.99 / 0.99 \\
			Age $>$ 75  &  1.43 / 1.59  &  0.26 / 0.22  &  0.92 / 0.99 \\
			Student  &  3.13 / 4.49  &  0.08 / 0.08  &  0.92 / 0.87 \\
			Unemployed  &  2.12 / 3.1  &  0.25 / 0.18  &  0.95 / 0.95 \\
			Employed  &  2.54 / 3.77  &  0.23 / 0.15  &  0.97 / 0.85 \\
			Homemaker  &  1.79 / 2.31  &  0.25 / 0.2  &  0.97 / 0.89 \\
			Retired  &  1.69 / 2.02  &  0.23 / 0.23  &  0.97 / 0.99 \\
			
			\hline
		\end{tabular}
	\end{center}
\end{table}

\subsection*{Effect of business category on the exponent $\gamma$}

Finally, we need to verify that the positive correlation between the median distance $\bar{d}$ and the amount of money spent $v$ does not depend on the type of purchases (i.e. business category) in the two provinces. The different business categories and their proportions of associated transactions are available in Table S\ref{TabS2}. The relationship between the distance traveled and the amount of money spent according to the business category is presented in Table S\ref{TabS3}. In most of the cases, the value of $\gamma$ is strictly higher than 0. Note that in some cases, like for the Restaurants business category for example, due to the presence of outlier (Figure S2) no log-log relationship between $\bar{d}$ and $v$ has been found.

\begin{table}
	\caption{\textbf{Percentage of transaction associated to each of the 20 business categories selected.}}
	\label{TabS2}
	\begin{center}
		\begin{tabular}{lcc}
			\hline
			\centering Category   & Barcelona &  Madrid             \\
			\hline
			
			Supermarket  &  18.13  &  16.1 \\
			Hypermarket  &  9.24  &  11.75 \\
			Gas Stations  &  12.18  &  11.06 \\
			Clothing store chain  &  4.35  &  7.23 \\
			Restaurants  &  8.38  &  6.59 \\
			Department store  &  2.12  &  5.08 \\
			Clothing store chain  &  5.54  &  4.58 \\
			Pharmacy, optical and orthopedics  &  4.23  &  3.78 \\
			Retail store  &  6.32  &  2.97 \\
			Hair and beauty  &  2.76  &  2.63 \\
			Fast food restaurants and chains  &  1.02  &  2.28 \\
			Bars and cafe  &  1.78  &  1.56 \\
			Shoe store  &  1.57  &  1.38 \\
			Toys and sports articles  &  1.52  &  1.37 \\
			Electronics, computers and appliances  &  1.49  &  1.35 \\
			Car dealership, garage and spare parts distributors  &  1.18  &  1.1 \\
			Bazaar  &  0.99  &  1.08 \\
			Bookshop, music shop and stationery  &  1.32  &  1.02 \\
			DIY store  &  0.75  &  0.95 \\
			Hospitals, clinics, doctors'  &  0.88  &  0.87 \\
			
			\hline
		\end{tabular}
	\end{center}
\end{table}

\begin{table}[!h]
	\caption{\textbf{Relationship between the distance traveled and the amount of money spent according to the business category.}}
	\label{TabS3}
	\vspace{0.5cm}
	\hspace{-1cm}
	\begin{tabular}{lccc}
		\hline
		\centering Category   & Median Distance (BCN / MAD)  & Slope (BCN / MAD) & R$^2$ (BCN / MAD)  \\
		\hline
		
		Supermarket  &  1.38 / 1.83  &  0.4 / 0.13  &  0.85 / 0.98 \\
		Hypermarket  &  2.08 / 2.4  &  0.37 / 0.25  &  0.91 / 0.93 \\
		Gas Stations  &  3.22 / 3.71  &  0.48 / 0.54  &  0.93 / 0.91 \\
		Clothing store chain  &  4.42 / 4.94  &  0.02 / -0.01  &  0.13 / 0.48 \\
		Restaurants  &  5.01 / 5.52  &  0.03 / -0.04  &  0.1 / 0.21 \\
		Department store  &  3.68 / 4.82  &  0.02 / -0.03  &  0.8 / 0.34 \\
		Clothing store chain  &  2.56 / 4.06  &  0.14 / 0.07  &  0.98 / 0.88 \\
		Pharmacy  &  1.52 / 2.01  &  0.15 / 0.02  &  0.89 / 0.27 \\
		Retail store  &  1.46 / 1.86  &  0.24 / 0.18  &  0.93 / 0.69 \\
		Hair and beauty  &  1.65 / 2.04  &  0.18 / 0.19  &  0.89 / 0.92 \\
		Fast food restaurants  &  4.94 / 5.31  &  -0.15 / -0.11  &  0.92 / 0.97 \\
		Bars and cafe  &  4.26 / 4.99  &  -0.07 / -0.11  &  0.34 / 0.89 \\
		Shoe store  &  2.21 / 3.36  &  0.21 / 0.12  &  0.98 / 0.92 \\
		Toys and sports articles  &  3.27 / 5.4  &  0.24 / 0.09  &  0.91 / 0.9 \\
		Electronics  &  4.56 / 5.96  &  0.07 / 0.1  &  0.93 / 0.89 \\
		Car dealership  &  3.35 / 4.64  &  -0.08 / -0.04  &  0.77 / 0.85 \\
		Bazaar  &  2.45 / 3.42  &  0.22 / 0.1  &  0.96 / 0.99 \\
		Bookshop  &  2.47 / 3.65  &  0.07 / -0.14  &  0.91 / 0.71 \\
		DIY store  &  5.3 / 8.93  &  0.15 / 0.1  &  0.98 / 0.92 \\
		Hospitals  &  2.62 / 3.5  &  0.08 / 0.06  &  0.9 / 0.45 \\
		
		\hline
	\end{tabular}
\end{table}

\begin{figure}[!ht]
	\centering 
	\includegraphics[width=12cm]{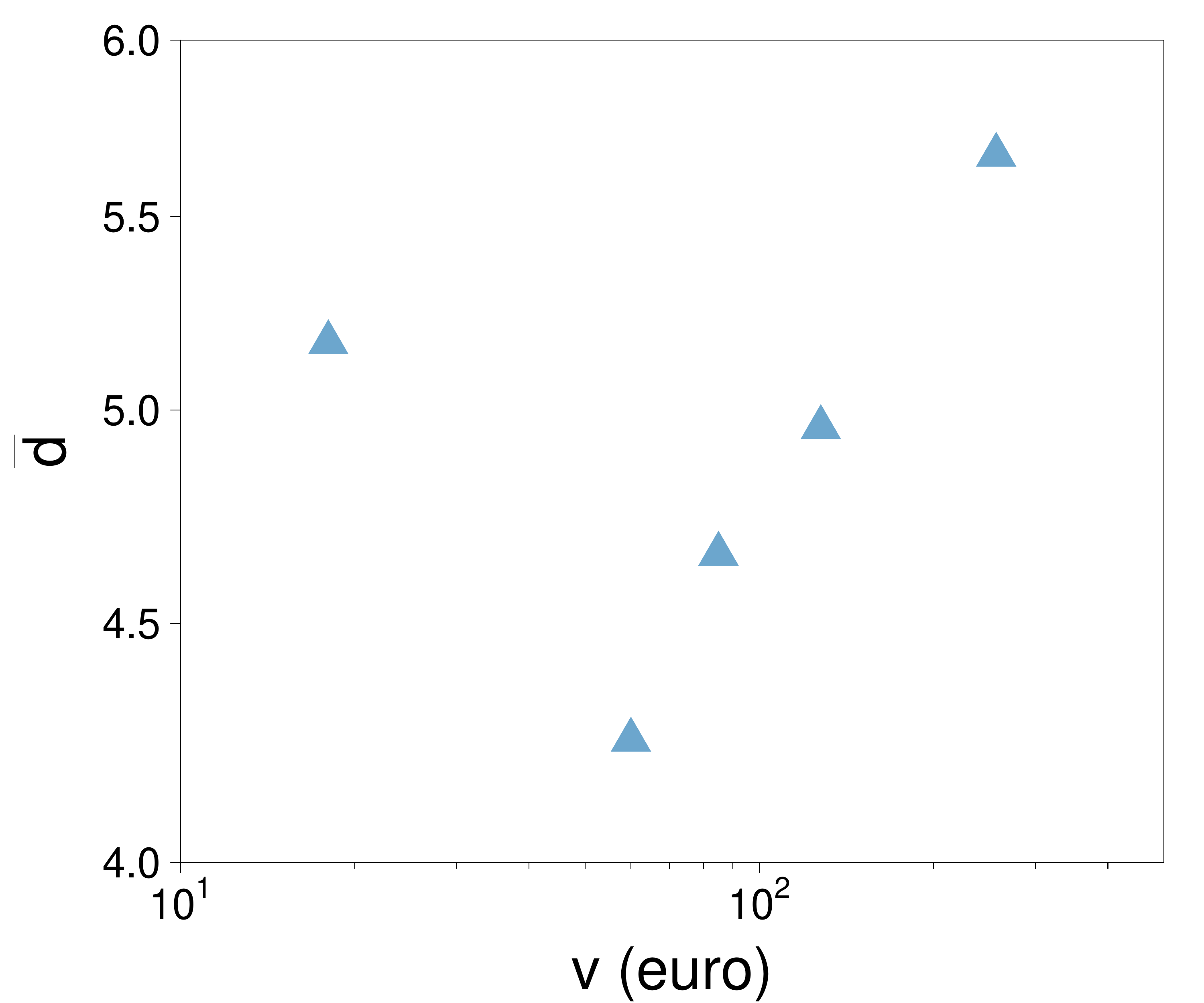}
	\caption{\textbf{Median distance $\pmb{\bar{d}}$ as a function of the amount of money spent in each bin for the Restaurants categorie.} \label{FigS2}}
\end{figure}

\clearpage
\newpage
\subsection*{Supplementary Figures}

\begin{figure*}[!ht]
	\centering
	\includegraphics[width=12cm]{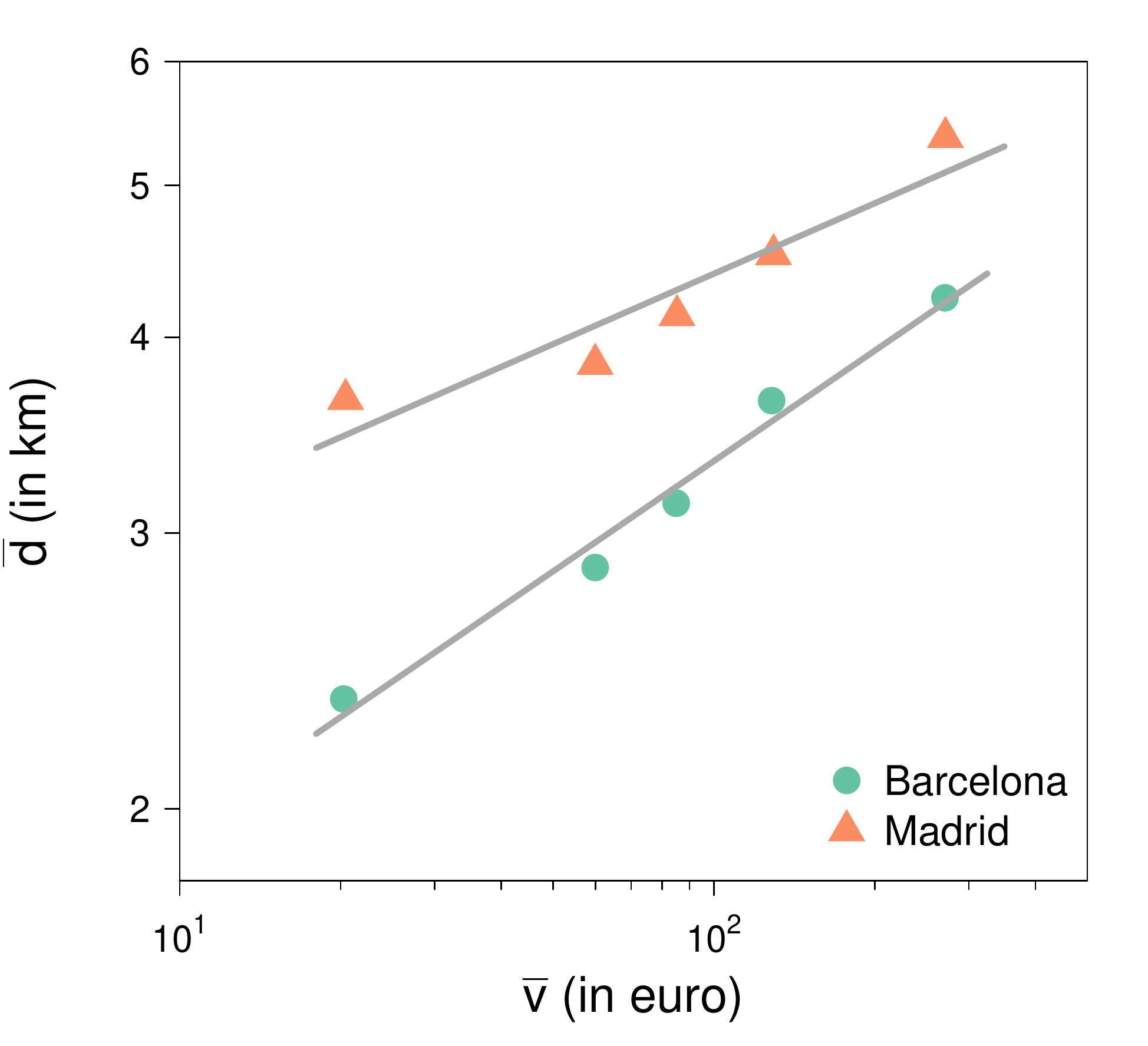}
	\caption{\textbf{Scaling relationship between amount of money spent and distance traveled based on daily-unique transaction.} Median distance $\bar{d}$ as a function of the median amount of money spent $\bar{v}$ for each range of prices in Barcelona (green dots) and Madrid (orange triangles). We define a daily-unique transaction as a transaction made by a user during a day where she or he made only one transaction. They represent 39.77\% of the transaction in Barcelona and 40.07\% in Madrid.\label{FigS3}}
\end{figure*}

\end{document}